\documentclass[12pt,letterpaper]{article}
\usepackage{jheppub}

\usepackage{color}

\usepackage{epsfig,multirow,subfigure} 
\usepackage{amscd}
\usepackage[matrix,arrow,curve]{xy}
\usepackage{verbatim}
\usepackage{latexsym}
\usepackage{amsfonts,amsthm,amsmath,mathtools}

\newcommand{\be}{\begin{equation}}
\newcommand{\ee}{\end{equation}}
\newcommand{\bea}{\begin{eqnarray}}
\newcommand{\eea}{\end{eqnarray}}
 
\title{\centering
A Note On 3D $\mathcal{N}=2$ Dualities:
\\
Real Mass Flow And Partition Function.
}

\author[a]{A. Amariti,}

\affiliation[a]{
  Laboratoire de Physique Th\'eorique de l'\'Ecole Normale Sup\'erieure and\\
  Istitute de Physique Th\'eorique Philippe Meyer, 24 Rue Lhomond,
  Paris 75005, France}

\emailAdd{amariti@lpt.ens.fr}

\abstract{We study two well-known classes of dualities in three
  dimensional $\mathcal{N}=2$ supersymmetric field theories. In the
  first class there are non trivial interactions involving monopole
  operators while in the second class the dual gauge theories have
  Chern-Simons terms in the action. An RG flows connecting the first
  dual pair to the second one has been studied in the past and tested
  on the partition function on the squashed three sphere.  Recently an
  opposite RG flow connecting the second dual pair to the first one
  has been studied in the case of unitary gauge groups. In this paper
  we study this flow on the partition function on the squashed three
  sphere. We verify that the equality between the partition functions
  of the original dual models is preserved in the IR, where the other
  dual pair is reached.  We generalize the analysis to the
  case of symplectic and of orthogonal groups.}

\begin{document}

\maketitle

\section{Introduction.}

At low energies asymptotically free quantum field theories can flow to
strong coupling, where a perturbative analysis is impossible.  A
complete non-perturbative analysis of these models is complicate and
some effective description is necessary.  One possibility consists of
finding a completely different theory that in the IR describes the
same degrees of freedom and correlators of the original one.  In such
a case the two models are dual in the IR.  Anyway it is not a trivial task
to find the dual model of a strongly coupled quantum field theory.

An useful laboratory in this search of dual models is supersymmetry.
Indeed many examples of dualities have been known for a long time in
four dimensions, like the Montonen-Olive duality in the maximal
supersymmetric case \cite{Montonen:1977sn}, the $\mathcal{N}=2$
Seiberg-Witten duality \cite{Seiberg:1994rs} and the $\mathcal{N}=1$
Seiberg duality of SQCD \cite{Seiberg:1994pq}. These four dimensional
dualities map the strongly coupled \emph{electric} regime of one
theory to the weakly coupled \emph{magnetic} regime of the dual theory.
Many extensions of these dualities have been studied in four
dimensions.

The search of analogous dualities in other dimensions is a natural
problem.  In the last years the three dimensional case attracted a great
interest.  For example a rich class of dualities exists when
$\mathcal{N}=2$.  This case shares the same number of supercharges,
four, as the $\mathcal{N}=1$ four dimensional case, and one may expect
some analogy  with Seiberg duality \footnote{In three dimensions there
  is another kind of duality, called mirror symmetry, that we 
will not discuss in this paper \cite{Intriligator:1996ex}.}.

Indeed it was shown in \cite{Aharony:1997gp} that a \emph{similar}
duality exists in three dimensions. This duality is usually called
Aharony duality.  This duality is only \emph{similar} to  Seiberg
duality because in the three dimensional case some differences arise
in the analysis of the moduli space and they affect the structure of
the dual field content and superpotential. More precisely, the
monopole operators that parameterize a branch of the moduli space of
one theory couple with the monopole operators of the dual theory.

There is a second class of dualities in the three dimensional case
that looks close to Seiberg duality.  In three dimensions
one can write a topological Chern-Simons (CS) action.  This action
modifies the vacuum structure of a theory and gives rise to the
Giveon-Kutasov duality \cite{Giveon:2008zn}.  This duality maps two
CS gauge theories and looks similar to the four dimensional
Seiberg duality in terms of the field content even if it involves CS
theories.  More recently other dualities that mix the Aharony and
Giveon-Kutasov cases have been found in \cite{Benini:2011mf}.  These
cases involve a chiral like matter content where the number of
fundamentals and anti-fundamentals does not coincide.

Even if Aharony and Giveon-Kutasov dualities look different, at
the level of the action and of the field content, they are strongly
related. The reason is that in three dimensions one can assign a real
mass to the matter fields charged under some global symmetries.  When
a gauge charged chiral fermion is massive it can be integrated
out. This process shifts the effective CS level by a semi-integer number.  This
real mass flow connects theories without CS terms to theories
with CS terms, and one can in principle flow from one pair of dual
theories to another thanks to this mechanism.  While the flow from the
Aharony pair to the Giveon-Kutasov is simple to understand the
opposite one is more mysterious, because one has to
understand the origin of the superpotential interaction between
the monopoles of the electric and of the magnetic phase.  This 
flow has been recently found  in \cite{Intriligator:2013lca} by observing
that if one assigns some large mass term to some of the electric
fields of the Giveon-Kutasov duality it does not simply reflects in
a set of masses for some dual fields. Indeed in the large mass regime
there are points in the moduli space that become singular, and one has
to map the vacua correctly.  Indeed at these points some massless
fields acquire a mass and some massive fields become
massless.  The flow investigated in \cite{Intriligator:2013lca} was
restricted to the case of CS level $k$ equals to $-1$. It was then
studied in some more general case in \cite{Khan:2013bba}.

In this paper we generalise the construction of \cite{Intriligator:2013lca}
to generic level $k$. Then we study the partition 
function on the squashed three sphere with the real mass terms.
We start from the equivalence of the partition
functions of the Giveon-Kutasov dual theories.
By adding the real masses on both sides
of this duality and by taking the large mass limit
we arrive at the expected relation for the Aharony dual pair.

An important aspect of the 
the computation is related to the structure of the
partition function. It is a matrix integral over the real scalar
component of the vector multiplet, reduced to the  Cartan subgroup.
In some phase this scalar may take a vev proportional to the large
mass that we introduced. Naively one may think that this shift cannot
affect the integral, but it is only true for a finite mass.  In the
large mass regime this vev affects the dominant contribution to the
partition function, such that the expected result is recovered.  We finally
study the analogous flow in the case of symplectic and orthogonal
groups, showing that similar results hold.

The paper is organised as follows.  In section \ref{reviewp} we review some
useful aspect of three dimensional $\mathcal{N}=2$ gauge theories that
are relevant in our analysis.  In section \ref{GKAd} we discuss
the Giveon-Kutasov and the Aharony dualities and we study the real
mass flow connecting them.  In section \ref{ZS3b} we review the matrix
integral describing the partition function on the squashed three
sphere and the role of the masses in the partition function.  In
section \ref{FFL} we study the RG flow from the Giveon-Kutasov to the
Aharony duality from the perspective of the partition function.  In
section \ref{SYMP} and \ref{ORTO} we study the symplectic and the orthogonal
cases respectively. In  section \ref{conc} we conclude.

\section{Review material.}
\label{reviewp}

In this section we review some basic aspects of $\mathcal{N}=2$ three
dimensional gauge theories. 
We refer the reader to \cite{Aharony:1997bx} for a more complete review.

Three dimensional $\mathcal{N}=2$ supersymmetry has four supercharges
$Q_{\alpha}$ and $\widetilde Q_\alpha$ with $\alpha=1,2$.  Their non
vanishing anticommutator is
\begin{equation}
\{ Q_{\alpha},\widetilde Q_{\beta}\} = \sigma_{\alpha \beta}^\mu P_\mu 
+ 2 i \epsilon_{\alpha \beta} Z
\end{equation}
where $Z$ is the central charge, corresponding to the reduced momentum
along $P_3$.

As in four dimensions there are a vector multiplet and a chiral (and
antichiral multiplet).  The vector multiplet $V$ is composed by a
gauge boson $A_\mu$, the gaugini $\lambda_\alpha$ and $\widetilde
\lambda_\alpha$ the $D$-term $D$ and a real scalar $\sigma$.  When
this real scalar gets an expectation value it generically breaks the
gauge symmetry to $U(1)^r$ ($r$ being the rank of the gauge group) and
one is at a generic point of the Coulomb branch.  In the abelian case
the photon can be dualized into a scalar $F_{\mu \nu,i} =
\epsilon_{\mu \nu \lambda} \partial^{\lambda} \phi_i$ ($i=1,\dots,r$)
and one can construct a supermultiplet $\Phi_i = \sigma_i + i \phi_i$.
This supermultiplet parameterizes the classical Coulomb branch and is
associated to a monopole operator $Y_i \simeq e^{\Phi_i}$.  The
Coulomb branch is usually lifted by quantum correction and only some
direction may remain flat.  There is also an Higgs branch,
parameterized by the charged chiral multiplets.

In three dimensions there are usually more global symmetries than in
the four dimensional case. Indeed the anomalous symmetries of the
four dimensional case become non anomalous in three dimensions.  In
addition we have the usual $U(1)_R$ charge that rotates the supercharges,
as in four dimensions.
Another new symmetry is the topological $U(1)_J$ symmetry, that is
generated by the current $J_\mu^{i} = \epsilon_{\mu \nu \lambda}
F^{\nu \lambda,i}$ and shifts the dual photon.  There are $r$ $U(1)_J$
currents, but at quantum level, where the Coulomb branch is lifted,
just one combination may be left.

Another multiplet that one can construct in three dimensions is the
linear multiplet. A linear multiplet, say $\Sigma$ is defined by
$\epsilon^{\alpha \beta} D_\alpha D_\beta \Sigma = \epsilon^{\alpha
  \beta} \overline D_\alpha \overline D_\beta \Sigma = 0$ and its
lowest component is a real scalar.  One can define a linear multiplet
for every globally conserved current.

The linear multiplet is useful to understand the relation between the
central charge and the real masses.  In three dimensions we can indeed
turn on a real mass $m$ for a chiral field $X$, if this last is
charged under a global symmetry.
\begin{equation}
\int d^4 \theta  X^{\dagger} e^{m \theta \overline \theta} X
\end{equation}
There is a linear multiplet that contains the global current under
which the chiral field $X$ is charged. The central charge is
associated to a background superfield and it is the scalar component
of this linear multiplet, in this case $Z=m$ .  In general $Z=\sum q_i
m_i$ where $m_i$ are background linear multiplets.

There is another contribution to the central charge coming from the
real FI parameter.  If one turns on a background vector multiplet
$V_b$ for the topological $U(1)_J$ one has can add a term $\int d^4
\theta V_b \Sigma$ that integrating by parts corresponds to $\int d^4
\theta V \Sigma_b$. The background linear multiplet $\Sigma_b$ is a FI
for the gauge multiplet $V$ and contributes to the central charge.

Lastly in three dimensions there is also a topological CS action
$k^{ij}\int d^4 \theta \Sigma_i V_j$, that is gauge invariant (under
large gauge transformations) if the CS level $k^{ij}$ is an integer.
It is important to observe that chiral fermions with a real mass and
CS levels are strongly connected. 
A massive fermion $\psi$ has in
general real mass $m_\psi = m + \sum_i q^i_{\psi} \sigma_i$.  By integrating
it out we have at one loop a shift in the CS level $(k_{eff})^{ij} =
k^{ij} + \frac{1}{2} \sum q_{\psi}^i q_{\psi}^i sgn(m)$.  By gauge
invariance $k_{eff}$ has to be integer. It implies that if $\sum
q_\psi^i q_\psi^j$ is odd and $k^{ij}$ is not vanishing then parity
is broken. This has to be compensated by $k^{ij} \in
\mathbb{Z}+\frac{1}{2}$. This phenomenon is named parity anomaly
\cite{Niemi:1983rq,Redlich:1983dv}.

\section{Giveon-Kutasov and Aharony  duality.}
\label{GKAd}

In this section we present the models that we investigate in the
rest of the paper. They are three dimensional $\mathcal{N}=2$
supersymmetric gauge theories with $U(N_c)$ gauge groups and $N_f$
matter fields in the fundamental and in the antifundamental
representation of the gauge group.  These theories are called vector
like because the number of fundamentals and of antifundamentals is the
same.  We leave possible generalisation to dualities between theories
with a chiral field content \cite{Benini:2011mf} for future
investigations.

First we discuss the Giveon-Kutasov duality \cite{Giveon:2008zn}.
The electric theory consists of a $U(N_c)$ gauge theory 
with a CS action at level $k$. There are $N_f$ 
fields $Q$ in the fundamental and $N_f$ fields $\widetilde Q$
in the antifundamental of the gauge group.
In absence of superpotential there is a $SU(N_f) \times SU(N_f)$
flavor symmetry acting on these quarks. Moreover we have an $U(1)_A$
global symmetry under which both the fields have the same charge $+1$
and an $U(1)_R$ symmetry 
%under which we assign to the field
%a generic $R$-charge $\Delta$ (the exact value of this charge is fixed by the 
%$F$-maximization \cite{Jafferis:2010un,Closset:2012vg})
.

The dual theory is a  $U(N_f-N_c+|k|)$ gauge theory 
with a CS action at level $-k$. There are $N_f$ 
fields $q$ in the fundamental and $N_f$ fields $\widetilde q$
in the antifundamental of the gauge group.
The $N_f^2$ electric mesons $M= Q \widetilde Q$
are elementary singlets in the dual description and couple to the
quarks through the superpotential
\begin{equation} \label{eq:spotGK}
W = M q \widetilde q
\end{equation}
The charges of the fields under the global symmetries are
\begin{equation}
\begin{array}{c||c|c|cccc}
  &U(N_c) &U(\widetilde N_c) & SU(N_f) & SU(N_f)& U(1)_A &U(1)_R \\
  \hline
  Q   & N_c  & 1 & \overline N_f &1  &1  &\Delta\\
  \tilde Q   & \overline N_c &1  &1  & N_f & 1 &\Delta\\
  \hline
  q   & 1 &\overline{\widetilde{N_c }} & N_f &1  & -1 & 1-\Delta\\
  \tilde q   & 1 & \widetilde{N_c } & 1 & \overline  N_f &  -1&1-\Delta\\
  M   & 1 & 1                        & \overline  N_f & N_f  & 2 &2 \Delta
\end{array}
\end{equation}
In this paper we are interested in connecting these two models with
another pair of dual models by an RG flow.  This second pair of dual
models, the Aharony duality, was found in \cite{Aharony:1997gp}.  On
the electric side we consider a $U(N_c)_0$ gauge theory, where the
subscript indicates that there is not CS action, and $N_f$
fundamentals and antifundamentals $Q$ and $\widetilde Q$ respectively.
The dual model has gauge group $U(N_f-N_c)_0$, the CS level is
vanishing as well, and there are $N_f$ fundamentals and
antifundamentals $q$ and $\widetilde q$ respectively.  As in the case
of the Giveon-Kutasov duality the electric mesons are elementary
degrees of freedom in the dual description and couple to the quarks
through the superpotential (\ref{eq:spotGK}).

The Coulomb branch is not
completely lifted by the quantum corrections. There are still 
combinations of monopole operators that remains flat, and they
correspond to the ones with flux $(\pm 1,0,\dots,0)$ in the Cartan of
the gauge group. We will refer to these monopole operators in the
electric theory as $X_{\pm}$.  They have charge $\pm 1$
under the topological $U(1)$ that shifts the dual photon.  In the
magnetic theory they are singlets that interact with the dual monopole
operators $x_{\pm}$  by a superpotential
\begin{equation}
\Delta W = x_+ X_-+x_-X_+
\end{equation}
The charges of the fields under the global symmetries are
\begin{equation}
\begin{array}{c||c|c|ccccc}
  &U(N_c) &U(\widetilde N_c) & SU(N_f) & SU(N_f)& U(1)_A &U(1)_R&U(1)_J \\
  \hline
  Q   & N_c  & 1 & \overline N_f &1  &1  &\Delta&0\\
  \tilde Q   & \overline N_c &1  &1  & N_f & 1 &\Delta&0\\
  \hline
  q   & 1 &\overline{\widetilde{N_c }} & N_f &1  & -1 & 1-\Delta&0\\
  \tilde q   & 1 & \widetilde{N_c } & 1 & \overline  N_f &  -1&1-\Delta&0\\
  M   & 1 & 1                        & \overline  N_f & N_f  & 2 &2 \Delta&0\\
  X_+ &1&1&1&1&-N_f&N_f(1-\Delta)-N_c+1&1\\
  X_-&1&1&1&1&-N_f&N_f(1-\Delta)-N_c+1&-1\\
  x_+&1&1&1&1&N_f&N_f(\Delta-1)+N_c+1&1\\
  x_-&1&1&1&1&N_f&N_f(\Delta-1)+N_c+1&-1
\end{array}
\end{equation}

\subsection{Flowing from Aharony to Giveon-Kutasov duality.}
\label{AGK}
In this section we review the RG flow connecting the Aharony dual pair
to the Giveon-Kutasov dual pair studied in \cite{Willett:2011gp}.  The
CS terms are generated by assigning real masses to some of the
fermions and by integrating them out.  

We consider, on the electric side, a $U(N_c)_0$ gauge theory with
$N_f+k$ fundamentals and antifundamentals.  We turn on positive real
masses for these matter fields such that there are $N_f$ light and $k$
heavy fields.  By integrating the heavy massive fermions out we shift
the CS level from $0$ to $k$.  We are left with a $U(N_c)_k$ model
with $N_f$ fundamentals and antifundamentals.

The dual theory has an $U(N_f+k-N_c)_0$ gauge group with $N_f$ light
and $k$ heavy flavors.  Moreover this theory has new elementary
singlets, consisting of mesons and monopoles of the electric
theory. The monopoles and some components of the meson acquire a large
mass term, in accordance to the global symmetries.  By integrating the
heavy fields out we are left with a $U(N_f-N_c+k)_{-k}$ gauge theory,
with $N_f$ light flavors and $N_f^2$ light singlets. It corresponds to
the expected Giveon-Kutasov dual phase.

Observe that we could also have inverted the sign of the real masses.
It would have changed the electric level from $k$ to $-k$ and the dual
CS level in the opposite way.  In any case the rank of the dual is
$(N_f-N_c+k)$, for positive $k$.  In general, the dual of
$U(N_c)_k$  is $U(N_f-N_c+|k|)_{-k}$ for any choice of sign of $k$.

\subsection{Flowing from Giveon-Kutasov to Aharony duality.}
\label{GKA}

In this section we discuss the RG flow connecting the Giveon-Kutasov
dual pair to the Aharony dual pair.  This flow has been recently
studied in \cite{Intriligator:2013lca}.  The analysis of
\cite{Intriligator:2013lca} is restricted to the electric
$U(N_c)_{-1}$ gauge theory with $N_f$ flavors, dual to
$U(N_f-N_c+1)_{1}$ with $N_f$ flavors and $N_f^2$ singlets.  The
analysis has been extended to the cases with $k=2$ and $k=4$ in
\cite{Khan:2013bba}.  Here we generalise part of the analysis of
\cite{Intriligator:2013lca,Khan:2013bba} to general $k$.

In the case of $k=-1$ two different flows are possible.  In one case
one has in the IR the usual Aharony dual pair.  This has been named
the $1-4$ duality in \cite{Intriligator:2013lca} and we will keep the
same name in this paper.  In the second case, named $2-3$, one flows
to a slightly different pair of dual theories. The matter content is
the same as in the usual case, but one of the interaction between the
electric and magnetic monopoles in the dual phase is lifted and it
appears in the electric superpotential. The electric theory has $W=
x_- X_+$, a monopole of the magnetic theory is a singlet in the
electric phase.  The dual superpotential is $W= M q \tilde q + x_+
X_-$. This duality can be obtained from Aharony duality by adding a
superpotential mass term for the electric monopole (or antimonopole) in
the dual theory.  If $|k|>1$ more complicate structures are possible
\cite{Khan:2013bba}. We will not discuss these possibilities here.

First we study the flow from the Giveon-Kutasov dual
pair at level $-k$ ($k>0$) to the $1-4$ Aharony dual pair. 
We consider an
$U(N_c)_{-k}$ gauge theory with $N_f$ light fundamentals and $k$
heavy ones, with real masses.
For simplicity we take all the heavy masses with the same value $m>0$.
After integrating out the heavy
matter we obtain an $U(N_c)_0$ gauge theory with $N_f$ massless
flavors.  The level shifts because we
integrate out $2k$ chiral fermions with the same positive real mass.
This is the electric version of the Aharony duality.

The dual theory is a $U(N_f-N_c+2 k)_{k}$ gauge theory with $N_f$
light and $k$ heavy flavors, with mass $-m$.  There are $(N_f+k)^2$
singlets $M$, with $N_f^2$ light components. This theory has
superpotential $W = M q \tilde q$. In presence of the large real mass
one has to shift the vacuum parameterized by $\sigma_i$.
The shifted location on the Coulomb branch is interesting because 
some new light degrees of freedom arise here.
The magnetic Giveon-Kutasov phase flows to the magnetic Aharony  phase
if one chooses
$\sigma_{1} =\dots, \sigma_{N_f-N_c}=0$ and $\sigma_{N_f-N_c+1},\dots,
\sigma_{N_f-N_c+k}=-\sigma_{N_f-N_c+k+1},\dots,-\sigma_{N_f-N_c+2k}=m$.
All the other matter fields are at the origin.

This non trivial vacuum is crucial in the analysis, because it
corresponds to a point in the moduli space where the gauge symmetry is
broken and some of the real masses for the quarks become light.  The
gauge group $U(N_f-N_c+2 k)_{k}$ is broken to $U(N_f-N_c) \times
U(k)^2$.  After integrating out the heavy quarks the $U(N_f-N_c)$
factor has $N_f$ light flavors and vanishing CS level.  The two $U(k)$
factors are more involved. Indeed in this case the light $N_f$ quarks
become heavy because their mass is shifted by the vev of $\sigma$.
The mass of the extra $k$ heavy fields is shifted in the two sectors
by an amount of $\pm m$.  In one case we are left with $k$ light
fundamentals and $k$ antifundamentals with mass $-2m$, in the other
case the situation is the opposite, there are $k$ light
antifundamentals and $k$ heavy fundamentals, with mass $-2m$.  In both
cases an effective CS level $\frac{k}{2}$ is obtained. Summarizing
the $U(N_f-N_c)_0$ sector has $N_f$ flavors and $N_f^2$ singlets, one
$U(k)_{k/2}$ sector hase $k$ charged chirals and the other
$U(k)_{k/2}$ has $k$ charged antichirals.

The last two sectors are the key ingredients to obtain the monopole
interactions in the Aharony duality.  We consider one of these sectors
but the discussion applies in the same way to the second one.  A
$U(k)_{\frac{k}{2}}$ gauge theory with one chiral superfield with
charge $+1$ is dual to a single chiral superfield $X_+$
\cite{Benini:2011mf} \footnote{One can understand this duality
  starting from Aharony duality with an $U(k)_0$ gauge group and $k$
  flavors. By adding a large positive (negative) real mass to the $k$ chiral or
  to the $k$ antichiral fields one generates a positive (negative) CS level. On
  the other hand, in the dual theory, that is composed just by
  singlets, many fields become massive and one remains with just one
  singlet.}.
This operator \emph{must} couple with the magnetic
monopoles $x_-$ associated to $U(1) \subset U(N_f-N_c)$.  The reason
is that the original $U(N_f-N_c+2k)$ theory has a topological symmetry
and under this symmetry the operators $X_+$ and $x_-$ have opposite
charges.  The coupling is through a superpotential $W = x_- X_+$
\cite{Intriligator:2013lca}.  In the second case we can repeat the
same analysis and we eventually obtain $W = x_+ X_-$.  Finally we get
a $U(N_f-N_C)_0$ with superpotential
\begin{equation} 
W = M q \tilde q + x_- X_+ + x_+ X_-
\end{equation}
as predicted by Aharony duality.  Observe that here we are restricting
to the case of level $-k$ in the electric theory with positive $k$.
The case of $k<0$ can be studied by inverting the sign of the real
mass.

As discussed in \cite{Intriligator:2013lca} there is a second
possibility. It consists of studying the electric theory in the non
trivial vacuum $\sigma_{N_c-k},\dots,\sigma_{k}=m$, and all the other
components vanishing.  The magnetic vacuum that preserves the duality
is $\sigma_{N_f-N_c+k+1},\dots,\sigma_{N_f-N_c+2k}=-m$, and all the
other components are vanishing.  In the large $m$ limit the electric
theory is a $U(N_c-k)_0 \times U(k)_{-k/2}$ gauge theory with $N_f$
light fundamentals and antifundamentals, charged under $U(N_c-k)_0 $,
and $k$ light fundamentals, charged under $U(k)_{-k/2}$.  The magnetic
theory is a $U(N_f-N_c+k)_0 \times U(k)_{k/2}$ gauge theory with $N_f$
light fundamentals and antifundamentals, charged under $U(N_c-k)_0 $,
and light $k$ fundamentals, charged under $U(k)_{k/2}$.  The two extra
$U(k)$ sectors can be studied as before and we obtain the electric
superpotential $W = x_- X_+$ and the magnetic one $W = M q \tilde q +
x_+ X_-$.  We refer to this duality as the $2-3$
duality.

\section{The squashed three sphere partition function.} 
\label{ZS3b}

In this section we review some aspect concerning the partition
function on a squashed three sphere, $S_b^3$,  preserving a $U(1)^2$
isometry of the original $SO(4)$ of the round case.

Partition functions computed on curved backgrounds that preserve some
supercharges are powerful objects, because they give one loop exact
results in supersymmetry.  Localization is the most general technique
to perform these calculations and it was first used in
\cite{Pestun:2007rz} for the partition function on $S^4$ of
$\mathcal{N}=2$ four dimensional theories.  The case of the three
dimensional sphere
was first studied in \cite{Kapustin:2009kz} for $\mathcal{N}>2$.  The
extension to $\mathcal{N}=2$ was done in
\cite{Jafferis:2010un,Hama:2010av} for the round sphere and in
\cite{Hama:2011ea} for the case we are interested in.

The possibility of computing the partition function of a quantum field
theory exactly is useful in checking the dualities.  Indeed many
dualities proposed in three dimensions have been checked by matching
their partition functions on $S^3_b$.

Moreover one can add mass contributions to the partition function and
study properties of the RG flows.  Indeed in this paper we are interested in
checking the flow by considering the partition functions of the
Giveon-Kutasov dual pair and by checking that, at the end of the
flow, the two partition functions still match and describe the correct
Aharony dual pair.

The general structure of the partition function on the squashed sphere
for a gauge group of rank $G$ and charged matter is 
\begin{eqnarray}
\label{eq:SquashedSphere}
Z = \frac{1}{|W|}\int
\prod_{i=1}^{G}\frac{ d\sigma_i}{\sqrt{-\omega_1 \omega_2}}
e^{\frac{i k \pi \sigma_i^2}{\omega_1 \omega_2}+\frac{ \pi i \lambda \sigma_i}{\omega_1 \omega_2}}
\frac{\prod_I \Gamma_h\left(
\omega \Delta_I + \rho_I(\sigma)+\widetilde \rho_I(\mu)\right)
}{\prod_{\alpha \in R_+}\Gamma_h\left(\alpha(\sigma)\right)
\Gamma_h\left(-\alpha(\sigma)\right)}
\end{eqnarray}
The integral is performed over the Cartan subgroup of the gauge
group. It is parameterized by the diagonal entries of the real scalar
$\sigma$ in the gauge group.  The exponential receives contributions
from the classical action, from the CS term at level $k$ and from the
real FI parameter $\lambda$.  $|W|$ represents the sum over the Weyl
degeneracies.

The Gamma $\Gamma_h$ functions are obtained by computing the one loop
superdeterminants of the vector and matter multiplets. They are
usually divergent expressions that require a regularization.  The
function $\Gamma_h$ is referred in the literature as hyperbolic Gamma
function and it can be written as
\begin{equation}
\label{eq:Gammah}
\Gamma_ h(z;\omega_1,\omega_2) \equiv \Gamma_h(z) \equiv
\prod_{n,m=1}^{\infty}
\frac{(n+1)\omega_1+(m+1) \omega_2-z }{n \omega_1+m \omega_2+z}
\end{equation}
The contribution of the vector multiplet corresponds to the denominator of
(\ref{eq:SquashedSphere}) and it is parameterized by the positive
roots of the algebra \footnote{Actually in the one loop determinant of
  the vector multiplet there is another term that cancels against the
  Vandermonde determinant in the measure.}.  The contribution of the
matter multiplet is the last term in the numerator of
(\ref{eq:SquashedSphere}).  Each term corresponds to the contribution of
the $I$-chiral multiplet with $R$ charge $\Delta$. The $I$-field is in
the representation $r$ of the gauge group $G$ with weight
$\rho_I(\sigma)$ and in the representation $\widetilde r$ of the
flavour group $F$, with weight $\rho_I(\mu)$. Sometimes in the rest of
the paper we will use the shortcut
$\Gamma_h(x)\Gamma_h(-x)=\Gamma_h(\pm x)$ to simplify the expressions.

In the rest of this section we discuss the relation between the masses and
the partition function, because it is crucial in the analysis of the
RG flow.  As observed, the partition function has an explicit
dependence on the Cartan subgroup of the flavor symmetry. This is related to
appearance of the central charge in the supersymmetry algebra.  The
central charge is $Z = \sum q_i m_i$ where the sum is performed over
the real masses of the matter fields and $q_i$ are the charges of
these fields under the flavor symmetries.

Observe that these global non-R currents usually mix with the $U(1)_R$
current.  This mixing is associated to the assignation of a non zero
imaginary part to the real masses.  The exact $R$-charge is obtained
by finding the combination that minimizes the absolute value of the
partition function \cite{Jafferis:2010un,Closset:2012vg}.

Observe that the monopole operators are charged also under the global
topological $U(1)_J$ that shifts the dual photon. It follows that this
charge mixes with the $R$-charge for these operators, and we can take
this mixing into account in the partition function by assigning an
imaginary part to the parameter $\lambda$.

One can consider another type of mass term. It consists of a complex
mass in the superpotential. Suppose we have a massive combination 
\begin{equation}
W = m Q \tilde Q
\end{equation}
where $Q$ and $\widetilde Q$ are in the  and anti-fundamental of the
gauge group $G$ and their $R$ charges are
$\Delta_Q=\Delta=2-\Delta_{\widetilde Q}$.  Thanks to the relation
$\Gamma_h(z)\Gamma_h(2\omega-z)=1$ the contribution of $Q$ and
$\widetilde Q$ to the partition function is $1$.

The relation of the partition function with the real masses is more
interesting. Indeed integrating out a real mass term modifies the
partition function.  Let us consider a field with $R$-charge
$\Delta_I$ charged under the gauge group with weight $\rho_I(\sigma)$
and real mass $\rho(\mu_I)+m$.  In the large $m$ limit $\Gamma_h$
reduces to
\begin{eqnarray}
\label{eq:intout}
&&\lim_{m \rightarrow \pm \infty}
\Gamma_h\left(\omega \Delta_I 
+ \rho_I(\sigma)
+\widetilde \rho_I(\mu)
+m\right)= \nonumber\\
=&&
\zeta^{-sgn(m)} e^{\frac{i \pi}{2\omega_1 \omega_2} sgn(m)
\left(\omega( \Delta_I-2) 
+ \rho_I(\sigma)
+\widetilde \rho_I(\mu)
+m\right)^2 }
\end{eqnarray}
We will often use this relation   in this paper.
Indeed as discussed in section \ref{GKA} the flow from 
the Giveon-Kutasov to the Aharony duality is performed by
assigning large real masses to some of the fields.

By using the relation (\ref{eq:intout}) we have to decouple the effect
of the massive fields on both sides of the duality. This is necessary
to match the partition functions of the new dual pair, where the heavy
fields disappear.  However the large mass dependence is not only in the matter
fields, because extra mass dependencies come from the vacuum
structure.  Indeed in some cases the scalar $\sigma$ in the vector
multiplet takes an expectation value.  In these cases one has to shift
$\sigma$ by an amount of $m$, where $m$ is the large mass.  This shift
affects the integral in the large $m$ limit, modifying the dominant
contribution.  We refer the reader to \cite{VdB,
  Niarchos:2012ah,Aharony:2013dha} for more discussions on this point.

Moreover, there are extra contributions of the real masses to the
partition function, related to CS contact terms associated to the
global symmetries \cite{Benini:2011mf,Closset:2012vp,Closset:2012vg}.
These contributions are necessary in the flows that we are considering
for the decoupling of the heavy masses on the two sides of the
duality.
 
We conclude this section by reviewing the relations between the two
sides of the Aharony and Giveon-Kutasov  dualities.  These relations have
discovered in \cite{VdB} and studied in the physical literature in
\cite {Benini:2011mf, Willett:2011gp, Niarchos:2012ah,
  Agarwal:2012wd,Aharony:2013dha}.
First we fix the gauge group $G=U(N_c)$ (we will review the
$O(N_c)$ and $SP(2N_c)$ case later). Then we consider $N_f^{(1)}$
quarks $Q$ and $N_f^{(2)}$ antiquarks $\tilde Q$.  We consider a real
mass $\mu_a$ for the fundamentals and $\nu_b$ for the
antifundamentals, where $a=1,\dots,N_f^{(1)}$ and
$b=1,\dots,N_f^{(b)}$.  From now on we absorb the $R$ charge inside
the real masses as well. This is done by turning on an imaginary
deformation (this can be done because the $U(1)_R$ mixes with the
abelian global symmetries).  In presence of a CS action, at level $k$
and of a real FI parameter $\lambda$ the partition function for such a
model is
\begin{equation}
\label{simpleCSU}
I_{U(N_c)_{k}}^{\left(N_f^{(1)},N_f^{(2)}\right)}(\mu;\nu;\lambda)
=\int \prod_{i=1}^{N_c} d \sigma_i
e^{\frac{ i \pi( k \sigma_i^2+ \lambda \sigma_i)}{\omega_1 \omega_2}}
\frac{\prod_{a=1}^{N_f^{(1)}}
\Gamma_h\left( \sigma_i  +\mu_a \right)\prod_{b=1}^{N_f^{(2)}}
\Gamma_h\left(-\sigma_i +\nu_b \right)}
{\prod_{1\leq i <j \leq N_c}\Gamma_h^{-1}\left(\pm(\sigma_i-\sigma_j)\right)}
\end{equation}
Now we consider the case of Aharony duality.
The electric side is obtained by turning off the CS level and 
by fixing  $N_f^{(1)}=N_f^{(2)}$.
The parameters $\mu_a$ and $\nu_b$ are
\begin{equation}
\mu_a = m_A + m_a + \omega \Delta
\quad \quad 
\nu_b = m_A + \widetilde m_b + \omega \Delta
\end{equation}
with the balancing condition
$\sum_{a=1}^{N_f}\mu_a=\sum_{b=1}^{N_f}\nu_b=m_A$.
The equivalence between the partition functions 
is encoded in the relation
\begin{eqnarray}
\label{eq:AZ}
I_{U(N_c)_{0}}^{\left(N_f,N_f\right)}(\mu;\nu;\lambda)
&=&
I_{U(N_f-N_c)_{0}}^{\left(N_f,N_f\right)}(\omega-\mu;\omega-\nu;-\lambda)
\nonumber 
\\
&\times&
\Gamma_h\left((N_f-N_c+1)\omega 
-N_f m_A \pm \frac{\lambda}{2}\right)
\prod_{a,b=1}^{N_f}\Gamma_h(\mu_a + \nu_b)
\end{eqnarray} 
The relation between the two phases of the Giveon-Kutasov
duality is given by (here we fix $k>0$)
\begin{eqnarray} \label{eq:GKZ}
&&
I_{U(N_c)_{k}}^{\left(N_f,N_f\right)}(\mu;\nu;\lambda)
=
I_{U(N_f-N_c+k)_{-k}}^{\left(N_f,N_f\right)}(\omega-\mu;\omega-\nu;-\lambda)
\prod_{a,b=1}^{N_f}\Gamma_h(\mu_a + \nu_b)
\zeta^{-k^2-2} e^{\frac{i \pi}{2\omega_1 \omega_2}\phi}
\nonumber \\
\end{eqnarray} 
where $\zeta=e^{\frac{i \pi}{24 \omega_1 \omega_2}}$ and
 the exponent $\phi$ is 
\begin{eqnarray}
\label{EXPO}
\phi &=& k(\sum_{a=1}^{N_f} (\mu_a^2+\nu_a^2)+k(k-2 (N_f-N_c+k))\omega^2
+\frac{1}{2}\lambda^2 - 2 k \omega \sum_{a=1}^{N_f} (\mu_a+\nu_a)
\nonumber \\
&+&
\lambda\sum_{a=1}^{N_f} (\mu_a-\nu_a)+
\frac{1}{2}(2(N_f-N_c+k)\omega- \sum_{a=1}^{N_f} (\mu_a+\nu_a))^2)
\end{eqnarray}
observe that here we always referred to $k>0$. The case $k<0$ 
is obtained from equation (5.5.7) of \cite{VdB}.

\section{Following the flow on the partition function.}
\label{FFL}

In this section we study the flow discussed in section \ref{GKA}.  We
first write the partition function of the dual Giveon-Kutasov pair,
by turning on the contribution of the real masses as well.  Then we
integrate out the heavy states by using formula (\ref{eq:intout}).
The decoupling of the large mass is triggered not only by the massive
fields but also by the vacuum structure.  Indeed, as already discussed,
the shift in the vev of sigma is not just a variable redefinition in
the large $m$ limit. If the vacuum is not at the origin an heavy quark
can be effectively light and the integral receives the
dominant contribution from such a vacuum.

\subsection{The 1-4 case.}

In this first case the vacuum of the
electric theory is $\langle
\sigma_1 \rangle = \dots=\langle \sigma_{N_c}\rangle = 0$ in the
electric case.  The gauge group is
$U(N_c)_{-k}$ with $k>0$ and $N_f$ light and $k$ heavy flavors.
The mass structure is
\begin{equation}
\label{MSE}
\left\{
\begin{array}{cccr}
\mu_a &=& m_a + m_A  \quad \quad \quad &a=1,\dots,N_f\\
\mu_a &=& m + m_A \quad\quad\quad& a=N_f+1,\dots,N_f+k\\
\nu_b &=& \widetilde m_b +m_A\quad\quad\quad&b=1,\dots,N_f\\
\nu_b &=&  m +m_A \quad\quad\quad&b=N_f+1,\dots,N_f+k
\end{array}
\right.
\end{equation}
In the large $m$ limit the partition function is
\begin{equation} \label{tatarahaele}
Z_{e} =  \lim_{m\rightarrow  \infty}
\zeta^{-2 k N_c}
e^{\frac{ i \pi k }{\omega_1 \omega_2} N_c(m+m_A-\omega)^2}
I_{U(N_c)_0}^{(N_f,N_f)}(\mu;\nu;\lambda)
\end{equation}
where we absorbed the $R$ charge inside the real masses.

The dual theory has gauge group $U(N_f-N_c+2k)_{k}$ with $N_f$ light
and $k$ heavy flavors.  
The mass structure is
\begin{equation}
\label{MSM}
\left\{
\begin{array}{cccr}
\mu_a &=& \omega-m_a - m_A  \quad \quad \quad &a=1,\dots,N_f\\
\mu_a &=& \omega-m - m_A \quad\quad\quad& a=N_f+1,\dots,N_f+k\\
\nu_b &=& \omega-\widetilde m_b -m_A\quad\quad\quad&b=1,\dots,N_f\\
\nu_b &=&  \omega-m -m_A \quad\quad\quad&b=N_f+1,\dots,N_f+k
\end{array}
\right.
\end{equation}
The mass structure
of the extra singlets $M$ can be read from the global charges of
$Q$ and $\widetilde Q$.  

The vacuum is given as in section \ref{GKA}. In the large $m$ limit
this changes the behaviour of the partition function and it is not a
simple shift of the variables.  This can be understood by writing the
explicit form of the partition function at large $m$.  In this first
example we separate each contribution coming from the vector
multiplet, the charged and the uncharged matter and the classical
action.

We start with the vector multiplet.  The non trivial expectation value
of $\sigma$ leaves three sectors, $U(N_f-N_c)\times U(k)^2$, and in
addition some mass dependent terms.  These mass dependent terms come
from the limit
\begin{eqnarray}
&&
\lim_{m\rightarrow \infty}
\prod_{i=1}^{N_f-N_c} \prod_{j=1}^{k} \Gamma_h(\pm (\sigma_i-\widetilde \sigma_j+m))
\Gamma_h(\pm (\sigma_i-\hat \sigma_j-m))
\times 
\prod_{i=1}^{k} \prod_{j=1}^{k}\Gamma_h(\pm (\widetilde\sigma_i-\hat \sigma_j-2m))
\nonumber \\
&&
=
\lim_{m\rightarrow \infty}
e^{\frac{4 i \pi  k m \omega  \left(N_f-N_c+k\right)}{\omega_1\omega_2}}
\prod_{i=1}^{k}e^{-\frac{2 i \pi  \omega (N_f-N_c+k) (\widetilde \sigma _i-\hat \sigma _i)}{\omega_1 \omega_2}}
\end{eqnarray}
The first term depends on the mass while the second term is a shift in the 
FI parameters of the unbroken $U(k)$ sectors.

In the charged matter sector there are three light contributions.
There are  $N_f$ fundamentals and antifundamentals  
in the unbroken $U(N_f-N_c)$,  
$k$ chiral fields in one of the unbroken $U(k)$ and
$k$ antichiral fields in the other unbroken $U(k)$.
The other charged fields are heavy and one has to integrate them out.
At large $m$ their contribution is
\footnotesize
\begin{eqnarray}
&&
\lim_{m\rightarrow \infty}
\prod_{i=1}^{N_f-N_c} \Gamma_h^k(\pm \sigma_i +\omega-m_A-m)
\prod_{i=1}^{k} 
\Gamma_h^k(\widetilde \sigma_i+\omega-m_A-2m)
\Gamma_h^k(-\hat \sigma_i+\omega-m_A-2m)
\nonumber \\
&&
\prod_{i=1}^{k} \prod_{a=1}^{N_f}
\Gamma_h(\widetilde \sigma_i -m +\omega-\mu_a)
\Gamma_h(-\widetilde \sigma_i +m +\omega-\nu_a)
\Gamma_h(\hat \sigma_i +m +\omega-\mu_a)
\Gamma_h(-\hat \sigma_i -m +\omega-\nu_a)
\nonumber \\
=&&
\zeta^{2k(N_f-N_c+k)}
\lim_{m\rightarrow \infty}
e^{\frac{i \pi k}{\omega_1 \omega_2}+m^2(N_f-N_c+4 k)-2 m m_A (N_c-3 N_f-2 k)+m_A^2 (N_f-N_c+k)}
\nonumber  \\
\times&&
\prod_{i=1}^{k}e^{\frac{i \pi  \left(m_A \left(2 N_f+k\right)+2 k m\right)}{\omega _1 \omega _2} (\widetilde \sigma_i-\hat \sigma_i)}
e^{-\frac{i \pi  k}{2 \omega_1+\omega_2}(\widetilde \sigma_i^2+\hat \sigma_i^2)}
\prod_{i=1}^{N_f-N_c}e^{-\frac{i \pi  k}{2 \omega_1+\omega_2}\sigma_i^2}
\end{eqnarray}
\normalsize
where we imposed the constraint 
$\sum_{a=1}^{N_f}(\mu_a+\nu_a)=2 m_A$.
This constraint, called balancy condition, has to be imposed 
on the masses in the case of Aharony duality.

The $(N_f+k)^2$ uncharged singlets split in three sectors.  One
contains light fields, i.e. the $N_f^2$ mesons of Aharony duality, the
other two sectors are heavy.  At large $m$ this heavy sector
contributes as
\begin{eqnarray}
&&
\lim_{m\rightarrow \infty}
\prod_{a=1}^{N_f} \Gamma_h^k(\mu_a+m_A+m) \Gamma_h^k(\nu_a+m_A+m)
\prod_{a,b=1}^{k} \Gamma_h(2m_A+2m)
\nonumber\\
=&&
\lim_{m\rightarrow \infty}
\zeta^{-k^2-2 k N_f}
e^{\frac{i \pi  k \left(\sum _{a=1}^{N_f} \left(\mu _a^2+\nu _a^2\right)+2 N_f \left(m_A+m-\omega \right) \left(3 m_A+m-\omega \right)+4 k m_A \left(m_A+2 m-\omega \right)+k (\omega -2 m)^2\right)}{2 \omega _1 \omega _2}}
\end{eqnarray}
The contributions from the classical action are shifted accordingly to
the vacuum structure.
The last contribution is the exponent (\ref{EXPO}).  By imposing the
constraint $\sum_{a=1}^{N_f}(\mu_a+\nu_a)=2 m_A$ it is
\begin{eqnarray}
&&
\zeta^{-k^2-2}e^{\frac{i \pi\left(-2 k \left(\sum _{a=1}^{N_f} (\mu _a^2+\nu _a^2)+2 k (m_A+m)^2\right)+4 \left(m_A (N_f+k)-\omega(N_f-N_c+2k)  N_c+k m \right)^2\right)}{4 \omega _1 \omega _2}}
\nonumber \\
\times
&&
e^{\frac{i \pi\left(8 k \omega  \left(m_A (N_f+k)+k m\right)-2 k \omega ^2 \left( -2 N_c+2 N_f+3 k \right)+\lambda ^2\right)}{4 \omega _1 \omega _2}}
\end{eqnarray}
\normalsize 
Finally at large $m$ the partition function of the dual
magnetic theory is
\begin{eqnarray} 
\label{largemdual}
Z_{m} &=& \lim_{m\rightarrow \infty}
e^{\frac{i \pi  \left(2 \omega  m_A \left(N_f \left(N_c-2 k\right)-N_f^2+k^2\right)+m_A^2 \left(k N_c+2 k N_f+N_f^2-2 k^2\right)+\omega ^2 
\left(N_c^2+N_f \left(N_f+2 k\right)-N_c \left(2 N_f+k\right)\right)\right)}{\omega _1 \omega _2}} 
\nonumber \\
&&
e^{\frac{i \pi  k m N_c \left(2 m_A+m-2 \omega \right)}{\omega _1 \omega _2}+\frac{i \pi  \lambda ^2}{4 \omega _1 \omega _2}}
~  \zeta^{-2 k  N_c-2}  ~
\prod_{a,b=1}^{N_f}\Gamma_h(\mu_a+\nu_b) ~
I_{U(N_f-N_c)_0}^{(N_f,N_f)}(\omega-\mu;\omega-\nu;-\lambda)
\nonumber \\
&&
I_{U(k)_{\frac{k}{2}}}^{(0,k)}(0;\omega-m_A;\lambda_1)~
I_{U(k)_{\frac{k}{2}}}^{(k,0)}(\omega-m_A;0;\lambda_2)
\end{eqnarray}
with
\begin{equation}\label{abelle}
\lambda_1 = 2 \omega (N_c-N_f)-m_A (k-2 N_f) - \lambda
\quad
,
\quad
\lambda_2 = -2 \omega (N_c-N_f) + m_A (k-2N_f) - \lambda
\end{equation}
The last line of (\ref{largemdual}) represents two chiral
$U(k)_{\frac{k}{2}}$ sectors, the first with $k$ fundamentals and the
second with $k$ antifundamentals.  In the field theory interpretation
each of these sectors is dual to a single chiral superfield.  These
integrals represent chiral CS gauge theories with level $\frac{k}{2}$,
for integer $k$. They have been studied in \cite{Benini:2011mf} and in
this case one can compute the integral.  We find
\begin{eqnarray}
I_{U(k)_{\frac{k}{2}}}^{(0,k)}(0;\omega-m_A;\lambda_1)&=&\zeta 
\Gamma_h\left(
\frac{k m_A}{2}-k \omega -\frac{\lambda _1}{2}+\omega
\right)
e^{\frac{i \pi  \left(2 k m_A (2 k \omega +3 \lambda_1 )+\lambda_1  (\lambda_1 -4 k \omega )-3 k^2 m_A^2\right)}{8 \omega _1 \omega _2}}
\nonumber \\
I_{U(k)_{\frac{k}{2}}}^{(k,0)}(\omega-m_A;0;\lambda_2) &=& \zeta 
\Gamma_h\left(
\frac{k m_A}{2}-k \omega +\frac{\lambda _2}{2}+\omega
\right)
e^{\frac{i \pi  \left(2 k m_A (2 k \omega -3 \lambda_2 )+\lambda_2  (\lambda_2+4 k \omega  )-3 k^2 m_A^2\right)}{8 \omega _1 \omega _2}}\nonumber
\\
\end{eqnarray}
By substituting the expression in (\ref{abelle}) to $\lambda_1$ and
$\lambda_2$ we obtain the expected monopole contribution with $N_c$
colours and $N_f$ flavors
\begin{equation}
\Gamma_h \left(\omega(N_f-N_c+1)- N_f m_A \pm \frac{\lambda}{2}\right)
\end{equation}
By putting everything together we reproduce 
the formula (\ref{eq:AZ}).

\subsection{The 2-3 case.}

In this section we consider the second possibility that we discussed in section
\ref{AGK}.
In this case there is  a different vacuum structure in both the
electric and magnetic case.
In the electric theory 
some of the $\sigma_i$ acquire a large expectation
value proportional to the real mass $m$.
The duality with the massive Giveon-Kutasov
magnetic theory is preserved by modifying
the vacuum of the magnetic theory appropriately, as discussed in 
section \ref{GKA}.

This different choice of vacua does not modify 
the structure of the $U(N_c)_0$ and $U(N_f-N_c)_0$
unbroken parts of the two Aharony dual gauge theories
but it changes the superpotential interactions.

We start by considering an electric CS gauge theory with
$U(N_c+k)_{-k}$ and with the same masses as (\ref{MSE}).  We fix $k>0$
and study the limit of large $m \rightarrow \infty$, with the vacuum
given by $\sigma_i=0$ for $i=1,\dots,N_c$ and $\sigma_i=-m$
$i=N_c+1,\dots,N_c+k$.

At large $m$ the partition function of the electric theory 
becomes
\begin{eqnarray}
  Z_e\!\!\! &=&\!\!\!\! \lim_{m\rightarrow \infty}
\zeta^{-k^2 - 2 k N_c}  
e^{-\frac{i \pi  k m \left(2 m_A \left(-N_c-N_f+k\right)-m N_c+2 \omega  N_f+\lambda \right)}{\omega _1 \omega _2}
-\frac{i \pi  k \left(k \omega^2-2 N_c \left(\omega -m_A\right)^2+k m_A \left(m_A-2 \omega \right)\right)}{2 \omega _1 \omega _2}
  }
  \nonumber \\
  &\times&
  I_{U(k)_{-\frac{k}{2}}}^{(0,k)}(0;m_A;\lambda_1)
  I_{U(N_f-N_c)_{0}}^{(N_f,N_f)}(\mu;\nu;\lambda)
\end{eqnarray}
with 
\begin{equation} \label{lambdauno}
\lambda_1=\lambda+2 \omega \left(N_f-N_c\right) +\omega -m_A (k+2 N_f)
\end{equation}
The gauge group of the dual phase is $U(N_f-N_c+k)_{k}$ and the masses of
the charged matter fields are given in (\ref{MSM}) .  There are also
$N_f^2$ light singlets, while the other components are
heavy.  The vacuum is $\sigma_i=0$ for $i=1,\dots,N_f-N_c$ and
$\sigma_i=m$ for $i=N_f+1,\dots,N_f+k$.  At large $m$ the dual
partition function becomes
\begin{eqnarray}
Z_m &=& \lim_{m\rightarrow \infty}
\zeta^{-k^2-2k N_c-2}  
e^{-\frac{i \pi  k m \left(2 m_A \left(-N_c-N_f+k\right)-m N_c+2 \omega  N_f+\lambda \right)}{\omega _1 \omega _2}}
\nonumber \\
&\times&
e^{\frac{i \pi  \left(8 \omega  m_A \left(N_f \left(N_c-N_f-2 k\right)+k^2\right)
+m_A^2 \left(4 k N_c+4 N_f \left(N_f+2 k\right)-10 k^2\right)
+4 \omega ^2 \left(N_c^2-N_c \left(2 N_f+k\right)
+N_f \left(N_f+2 k\right)\right)+\lambda ^2\right)}{4 \omega _1 \omega _2}}
\nonumber \\
&\times&
\prod_{a,b}\Gamma_h(\mu_a+\nu_b)
I_{U(N_f-N_c)_{0}}^{(N_f,N_f)}(\omega-\mu;\omega-\nu;-\lambda)
I_{U(k)_{\frac{k}{2}}}^{(0,k)}(0;\omega-m_A;\lambda_2)
\end{eqnarray}
with 
\begin{equation} \label{lambdatilde}
\widetilde \lambda=-\lambda_2
+2 \omega \left(N_f-N_c\right) -m_A (k+2N_f)
\end{equation}
By computing the two integrals of the two chiral sectors
one obtains the correct contributions from the monopoles in 
the electric and magnetic sectors, and the correct result is found.

\section{The symplectic group.}
\label{SYMP}
In this section we extend our results to the case of the symplectic gauge
group $SP(2 N_c)$. First we discuss the Aharony and
Giveon-Kutasov dualities for symplectic groups then we review some
useful result for the partition function.  Then we will discuss the
flow from the Giveon-Kutasov dual pair to the Aharony dual pair.
Finally we test the result on the partition function as in the
unitarity case.

\subsection{Aharony and Giveon-Kutasov dualities.}

First we discuss the case of Aharony duality.
The electric theory is a $SP(2 N_c)_0$ gauge theory with $2N_f$ flavors $Q$.
Note that the group is real here and the field content is chiral like.

The magnetic theory is a $SP(2(N_f-N_c-1))$ gauge theory with $N_f$
flavors $q$ and a singlet in the $2N_f(N_f-1)$ representation of the
$SU(2 N_f)$ flavor symmetry.  In addition the monopole operator $Y$ of
the electric theory is a singlet in the dual theory and couples to the
magnetic monopole $y$.
The superpotential is
\begin{equation}
W = M q q + Y y
\end{equation}
The symplectic version of Giveon-Kutasov duality was first discussed
in \cite{Willett:2011gp}.  In this case the electric theory has
$SP(2N_c)_{2k}$ gauge theory with $N_f$ light flavors.  The factor of
$2$ in front of the CS level is related to the normalization of the
generators of the Lie algebra \cite{Willett:2011gp}.  The dual theory
is a $SP(2(N_f-N_c-1+|k|)$ gauge theory with $N_f$ flavors and the
meson in the $N_f(2N_f-1)$ representation, with superpotential
\begin{equation}
W = M q q
\end{equation}

\subsection{Partition functions.}

As in the unitary case also in the symplectic case the identities
relating these two dualities have been first derived in \cite{VdB}.
Here we explicitly show these relations.  

First we give the general expression for the 
 partition function of an $SP(2N_c)_{2k}$ gauge theory with $2 N _f$ fundamental
\begin{equation}
I_{SP(2N_c)_{2k}}^{2N_f}(\mu)=
\frac{1}{2^{N_c} N_c!}
\int \prod_{i=1}^{N_c} \frac{d\sigma_i e^{\frac{2 i \pi k }{\omega_1 \omega_2}\sigma_i^2}}
{\sqrt{-\omega_1 \omega_2}}
\prod_{a=1}^{2N_f} \Gamma_h (\pm \sigma_i +\mu_a)\Gamma_h^{-1}(\pm 2\sigma_i)
\!\!\!\!\!\!
\prod_{1\leq i<j\leq N_c} \!\!\!\!
\Gamma_h^{-1}(\pm \sigma_i \pm \sigma_j)
\end{equation}
When the CS term is vanishing we have the Aharony dual pair. In
this case the equivalence between the electric and magnetic partition
functions is encoded in the relation 
\begin{eqnarray}
\label{ASP}
&&
I_{SP(2 N_c)_0}^{2 N_f}(\mu)= I_{SP(2(N_f-N_c-1))_0}^{2 N_f} 
\Gamma_{h}(2(N_f-N_c)\omega-2 N_f m_A)
\prod_{1\leq a <b \leq 2 N_f} \Gamma_h(\mu_a+\mu_b)\nonumber \\ 
&&
\end{eqnarray}
with the balancy condition $\sum_{a=1}^{2 N_f} \mu_a = 2 N_f m_A$.
The second relation can be derived from (\ref{ASP}) by integrating out some
massive quarks.
The duality between the electric $SP(2N_c)_{2k}$ and the magnetic 
$SP(2(N_f-N_c+|k|-1)_{-2k}$ CS gauge theories is 
summarised in the relation
\begin{eqnarray}
\label{GKSP}
&&
I_{SP(2 N_c)_{2k}}^{2 N_f}(\mu)= I_{SP(2(N_f-N_c-1))_{-2k}}^{2 N_f} 
\prod_{1\leq a <b \leq 2 N_f} \Gamma_h(\mu_a+\mu_b)
\zeta^{sign(k)(|k|-1)(2|k|-1)}
\nonumber \\ 
&&
e^{\frac{i \pi}{2 \omega_1 \omega_2}\left(
- 4 N_c k \omega^2 - sign(k)   \left((2N_f(1-\Delta)-2N_c+1)\omega-\sum_{a=1}^{2 N_f} \mu_a \right)^2
+2 k \sum_{a=1}^{N_f} (\mu_a+\omega(1-\Delta))^2
+k(2 |k|-1)\omega^2\right)}
 \nonumber \\
\end{eqnarray}

\subsection{Flowing from Giveon-Kutasov duality to Aharony duality.}
\label{flowSP}

We consider a $SP(2 N_c)_{-2k}$ gauge theory with 
$2 N_f$ light and $2k$ heavy flavors. 
The masses are
\begin{equation}
\left\{
\begin{array}{cccr}
\mu_a &=& m_A+m_a \quad\quad &a=1,\dots,2 N_f  \\
\mu_a &=& m_A+m \quad\quad &a=2N_f+1,\dots,2 (N_f+k) 
\end{array}
\right.
\end{equation}
 with
$m \rightarrow  \infty$.
This theory flows to a $SP(2N_c)_0$ gauge theory with $2N_f$ light
fundamentals.
The dual theory has $SP(2(N_f-N_c-1+2k))_{2k}$ gauge group and
$2(N_f+k)$ fundamentals and $2(N_f+k)^2-(N_f+k)$ singlets. The masses 
can be read from the global symmetries as usual.
The duality is preserved if the vacuum is chosen as 
\begin{equation}
\label{vacuumSP}
\left\{
\begin{array}{cccr}
\sigma_i &=& 0& \quad \quad   i=1,\dots,N_f-N_c-1 \\
\sigma_i &=& m& \quad \quad   i=N_f-N_c,\dots,N_f-N_c+2k-1
\end{array}
\right.
\end{equation}
This breaks the dual gauge group into $SP(2(N_f-N_c-1)_0 \times
U(2k)_{k}$.  By integrating out the massive matter we have, in the
unbroken symplectic sector, $2N_f$ light quarks while in the unitary
sector we are left  with $k$ light fundamentals.  This chiral theory can
be studied as in the case of the unitary groups and it gives raise to
the term $y Y$ required by the Aharony duality.  In the rest of this
section we will test the validity of this RG flow on the partition
function.

\subsection{Following the flow on the partition function. }

In this section we study the 
flow just explained on the partition function.
At large $m$ the partition function of the electric theory  is
\begin{eqnarray}
Z_{E} =
\lim_{m \rightarrow \infty }
\zeta^{-4 N_c k}
e^{\frac{2 i \pi  k N_c (m_A+m-\omega)^2}{\omega _1 \omega _2}}
I_{SP(2 N_c)_0}^{2 N_f}(\mu)
\end{eqnarray}
The dual partition function in the large $m$ limit is
\footnote{As in the unitary case we impose the balancing condition
on the light fields $\sum_{a=1}^{2N_f} \mu_a = 2 N_f m_A$.}
\begin{eqnarray}
Z_M \!=\!\lim_{m\rightarrow \infty}
\zeta^{-4 N_c k -1}   e^{\frac{i \pi}{2\omega_1 \omega_2}\phi}
\!\!\!\!\!\!\!\!
\prod_{1\leq a <b \leq 2 N_f} 
\!\!\!\!\!\!\!\!
\Gamma_h(\mu_a+\mu_b)
%\nonumber \\
%&&
I_{SP(2 (N_f-N_c-1))_0}^{2 N_f}(\omega-\mu)
I_{U(2k)_k}^{(0,2k)}(0;\omega-m_A;\lambda)  
\nonumber \\
\end{eqnarray}
where the exponent $\phi$ is
\footnotesize
\begin{eqnarray}
\phi&=&
8 k \omega  m_A N_c+8 k m m_A N_c-16 k^2 \omega  m_A+4 k^2 m_A^2+8 k \omega  m_A+4 k m^2 N_c\nonumber \\
&-&
8 k m \omega  N_c-12 k \omega ^2 N_c+4 \omega ^2 N_c+12 k^2 \omega ^2-8 k \omega ^2+\omega ^2
\nonumber \\
&+&
8 \omega  m_A N_c N_f+4 k m_A^2 N_c-32 k \omega  m_A N_f+16 k m_A^2 N_f-8 \omega  m_A N_f^2\nonumber \\
&+&
4 \omega  m_A N_f+4 m_A^2 N_f^2-8 \omega ^2 N_c N_f+4 \omega ^2 N_c^2+16 k \omega ^2 N_f+4 \omega ^2 N_f^2-4 \omega ^2 N_f
\end{eqnarray}
\normalsize
and the complex FI term in the $U(2k)_{k}$ sector is 
\begin{equation}
\label{lambdaSP}
\lambda =4(N_f-N_c+k)\omega-2\omega -2 m_A(k+2 N_f)
\end{equation}
This integral can be computed explicitly ant it gives
\begin{equation}
\label{finSP}
Z_{U(2k)_k} = 
e^{-\frac{i \pi  \left(4 k m_A (4 k \omega -3 \lambda )+\lambda  (8 k \omega +\lambda )-12 k^2 m_A^2\right)}{8 \omega _1 \omega _2}}
\Gamma_h\left(
\omega+k m_A-2 k \omega +\frac{\lambda }{2}\right)
\end{equation}
After substituting (\ref{lambdaSP}) into (\ref{finSP}) 
the argument of $\Gamma_h$ is $2\omega(N_f-N_c)-2 m_A N_f$
that represents the correct contribution of the electric monopole
appearing as an elementary field in the magnetic phase of Aharony duality.
Finally by eliminating the $m$ dependence on both sides we recover
formula (\ref{ASP}) as expected.

\section{The orthogonal case.}
\label{ORTO}
In this section we repeat the analysis for the orthogonal groups.
Giveon-Kutasov and Aharony dualities have been studied in \cite{Benini:2011mf,Kapustin:2011gh,Hwang:2011ht,Aharony:2013kma}.

First we analyse the Giveon-Kutasov theory.  We consider a
$O(N_c)_{-k}$ gauge theory with $N_f$ light and $k$ heavy 
flavors. The real masses are
\begin{equation}
\left\{
\begin{array}{cccr}
\mu_a &=& m_A + m_a& \quad\quad a=1,\dots,N_f \\
\mu_a &=& m_A + m &\quad\quad a=N_f+1,\dots,N_f+k \\
\end{array}
\right.
\end{equation}
with large positive $m$.  The dual theory is a $O(N_f-N_c+2+2k)_{k}$
gauge theory with $N_f$ light and $k$ heavy
flavor and $\frac{(N_f+k)(N_f+k+1)}{2}$ singlets.  Their
mass structure is obtained from the global charges as usual. 
The vacuum of this dual theory is 
\begin{equation}\label{shiftOn}
\left\{
\begin{array}{cccr}
  \sigma_i &=& 0  \quad \quad &i=1,\dots,N_f-N_c+2\\
  \sigma_i &=&-m  \quad \quad & i=N_f-N_c+3,\dots, N_f-N_c+2+2 k
\end{array}
\right.
\end{equation}
This flow generates the Aharony dual pair for the orthogonal group.

As in the unitary and symplectic cases we study this flow on the
partition function.  However, in the orthogonal case, some subtleties
arise because the rank of the gauge group can be even or odd, and
this modifies some terms in the partition function.  The partition
function for a model with an $O(N_c)$ gauge group (with $N_c=2N$
or $N_c=2N+1$)  with $N_f$ flavors is
given by the following two relations
\small
\begin{eqnarray}
I_{O(2N)_{k}}^{n_f}(\mu)&=& \frac{1}{2^{N-1} N!}
\int \prod_{i=1}^{N} \frac{d\sigma_i e^{\frac{i \pi k}{\omega_1 \omega_2}\sigma_i^2}}{\sqrt{- \omega_1 \omega_2}} 
\prod_{a=1}^{N_f} \Gamma_h (\pm \sigma_i + \mu_a)
\prod_{1\leq i < j \leq n} \Gamma_h^{-1}(\pm \sigma_i \pm \sigma_j)
 \\
I_{O(2N+1)_{k}}^{n_f}(\mu)&=& 
\frac{\prod_{a=1}^{N_f} \Gamma_h(\mu_a)}{2^N N!}
\int \prod_{i=1}^{N} \frac{d\sigma_i e^{\frac{i \pi k}{\omega_1 \omega_2}\sigma_i^2}}
{\sqrt{- \omega_1 \omega_2}} 
\prod_{a=1}^{N_f} \Gamma_h (\pm \sigma_i + \mu_a)
\Gamma_h^{-1}(\pm \sigma_i) \!\!\!\!
\prod_{1\leq i < j \leq N} 
\!\!\!\!
\Gamma_h^{-1}(\pm \sigma_i \pm \sigma_j)
\nonumber
\end{eqnarray}
\normalsize
Even if the two expressions look different it has been shown in 
\cite{Benini:2011mf}
that one can write a single formula to match the cases of 
the Aharony and Giveon-Kutasov duality.
For the Aharony duality one has
\begin{equation}
\label{eq:ZonAharonyO(n)}
I_{O(N_c)_0}^{N_f} = I_{O(N_f-N_c+2)_0}^{N_f}
\Gamma_h\left(\omega (N_f-N_c+2)-\sum_{a=1}^{N_f} \mu_a\right)
\prod_{a,b=1}^{N_f} \Gamma_h(\mu_a+\mu_b)
\end{equation}
with the constraint $\sum_{a=1}^{N_f}\mu_a = N_f m_A$.
In any case one has to distinguish in the explicit calculations
if the gauge group has rank even or odd.
The relation (\ref{eq:ZonAharonyO(n)})
allows to derive the analogous equality for the Giveon-Kutasov duality.
This was done in  \cite{Benini:2011mf} and one has
\begin{eqnarray}
\label{eq:ZonGKO(n)}
I_{O(N_c)_k}^{N_f}
&=& I_{O(N_f-N_c+2+2 |k|)_{-k}}^{N_f}
\zeta^{\frac{sgn(k)(k|+1)(|k|+2)}{2}}
\prod_{a,b=1}^{N_f} \Gamma_h(\mu_a+\mu_b)
\nonumber \\
&\times&
e^{\frac{i \pi}{2\omega_1\omega_2}
\left(
-N_c k \omega^2 -sgn(k)\left((N_f-N_c+2+2|k|)\omega
-\sum_{a=1}^{N_f} \mu_a\right)^2
+k\sum_{a=1}^{N_f} \left(\mu_a-\omega\right)^2
+\frac{k}{2}(|k|+1)\right)
}
\nonumber\\
\end{eqnarray}
In the rest of this section we study the flow explained above and show
that it connects (\ref{eq:ZonGKO(n)}) to (\ref{eq:ZonAharonyO(n)}).

As anticipated the explicit calculation requires the knowledge of the parity of
the rank of the gauge group in the electric and in the magnetic theory.
In general there are eight possibilities, depending on the parity of $N_c$,
$N_f$ and $k$.
By explicit computations it is possible to observe that they 
reduce to the same formula.
In general starting from the $O(N_c)_k$ theory 
with $N_f+k$ flavors, $N_f$ light and $k$ heavy
as usual one arrives to an $O(N_c)_0$ model with
$N_f$ light flavors.
In the large $m$ limit we have
\begin{eqnarray}
\label{eleO}
Z_E &=& \lim_{m\rightarrow \infty} \zeta^{-2 k N_c} I_{O(N_c)_0}^{N_f}(\mu)
e^{\frac{i \pi  k N_c \left(m_A+m-\omega \right)^2}{\omega _1 \omega _2}}
\end{eqnarray} 
In the dual magnetic case, we assign the proper masses to the
fields according to the global symmetries. Then
we  shift $\sigma$ as in (\ref{shiftOn}).
After fixing $\sum_{a=1}^{N_f} \mu_a = N_f m_A$ in the large $m$ limit we have
\begin{eqnarray}
\label{dualO}
Z_M &=& \lim_{m\rightarrow \infty} \zeta^{-2 k N_c-1} 
 \lim_{m\rightarrow \infty} e^{\frac{ i \pi k}{\omega_1 \omega_2} \phi}
\prod_{a,b=1}^{n_f} \Gamma_h(\mu_a+\mu_b)
I_{O(N_f-N_c+2)_0}^{N_f}(\omega-\mu)
I_{U(k)_{\frac{k}{2}}}^{(k,0)}(m_A;0,\lambda)
\nonumber \\
\end{eqnarray}
where the exponent $\phi$ is
\begin{eqnarray}
\phi&=&
m_A^2 \left(k N_c+4 k N_f+N_f^2+k^2\right)+N_c \left(-2 \omega ^2 N_f+k \left(m^2-2 m \omega -3 \omega ^2\right)-2 \omega ^2\right)
\nonumber\\
&+&
\omega ^2 ((4 k+2) N_f+N_f^2+3 k^2+4 k+1)-2 m_A (\omega  ((4 k+1) N_f+N_f^2+2 k (k+1))
\nonumber\\
&+&
\omega ^2 N_c^2-N_c (\omega  N_f+k (m+\omega )))
\end{eqnarray}
and the complex FI parameter is
\begin{equation}
\lambda = 
2 \omega  \left(N_f-N_c+k+1\right)-m_A \left(2 N_f+k\right)
\end{equation}
The integral 
$I_{U(k)_{\frac{k}{2}}}^{(k,0)}(m_A;0,\lambda)$
can be computed with the usual technique of \cite{Benini:2011mf}. It gives 
\begin{equation} \label{ipergO}
I_{U(k)_{k/2}}^{(k,0)}(m_A;0,\lambda) =
e^{-\frac{i \pi  \left(-3 k^2 m_A^2+2 k m_A (2 k \omega -3 \lambda )+\lambda  (4 k \omega +\lambda )\right)}{8 \omega _1 \omega _2}}
\Gamma_h\left(\frac{k m_A}{2}-k \omega +\frac{\lambda }{2}+\omega\right)
\end{equation}
After substituting $\lambda$ all the exponents cancel when one equates
(\ref{eleO}) to (\ref{dualO}).  The hyperbolic $\Gamma$ function in
(\ref{ipergO}) becomes
\begin{equation}
\Gamma_h(\omega  \left(N_f-N_c+2\right)-m_A N_f)
\end{equation}
and as expected it coincides with the expected contribution
of the electric  monopole.

\section{Conclusions.}
\label{conc}

In this paper we studied RG flow connecting two sets
of dual pairs in three dimensions.
We showed that the matching between the two
partition functions in the UV pair is preserved by the flow once
the IR  pair is reached.
The interesting fact is that one can reconstruct from the partition
function the contribution of the monopole sector in the magnetic
Aharony phase.  We have also shown that the analysis holds in the
symplectic and in the orthogonal cases.

Some generalization of our work are possible.
First one may consider the flows studied in \cite{Khan:2013bba}. 
Indeed it may happen in these cases that some of the
extra sectors, that we used to reconstruct the monopole contributions,
decouple. Studying these flows with the partition function should be
useful for an understanding of this decoupling.  Another possibility
consists of flowing from the Giveon-Kutasov case to the dual pairs
with a chiral field content studied in \cite{Benini:2011mf}.  One may
also study real mass flows from UV duals with adjoint matter
\cite{Niarchos:2008jb,Niarchos:2009aa} or more complicate
representations and gauge groups \cite{Kim:2013cma,Park:2013wta} and
obtain IR dualities, and check the behaviour of these flows on the
partition function.

Another interesting aspect that we did not address in this paper is
the role of accidental symmetries. Indeed already at the level of the
Aharony dual there are critical values of the levels and ranks at
which some singlet become free \cite{Safdi:2012re}. This is expected
also for theories with a representations different than the
fundamental \cite{Morita:2011cs,Kapustin:2011vz,Agarwal:2012wd}.  It
would be interesting to understand if and how accidental symmetries
may be an obstruction for these flow.

\section*{Acknowledgements}
We thank F.~Benini, S.~Cremonesi, L.~Girardello, C.~Klare, and
A.~Zaffaroni for comments.  We thank K.~Intriligator for
discussions and comments on the draft.  We are also grateful to the
organisers of the Modave summer school, where part of the results of
this paper have been presented.

%\bibliographystyle{JHEP}
%\bibliography{bpsref}

\end{document}